\documentclass[prd,aps,a4paper,eqsecnum,floatfix,twocolumn]{revtex4}  

\usepackage{graphicx} 
\usepackage{amsmath,amsfonts,amssymb}
\usepackage{tikz}
\usepackage{appendix}
\usepackage[caption=false,justification=justified]{subfig}

\newcommand{\beq}{\begin{equation}}
\newcommand{\eeq}{\end{equation}}
\newcommand{\bea}{\begin{eqnarray}}
\newcommand{\eea}{\end{eqnarray}}
\newcommand{\be}{\begin{equation}}      
\newcommand{\ee}{\end{equation}}

\def\nn{\nonumber}

\begin{document}

\title{Scattering angle in a Topological Star spacetime: a self-force approach}

\author{Massimo Bianchi$^{1}$, Donato Bini$^{2}$, Giorgio Di Russo$^{3}$}
  \affiliation{
$^1$Dipartimento di Fisica, Universit\`a di Roma \lq\lq Tor Vergata" and Sezione INFN Roma2, Via della
Ricerca Scientifica 1, 00133, Roma, Italy\\
$^2$Istituto per le Applicazioni del Calcolo ``M. Picone,'' CNR, I-00185 Rome, Italy\\
$^3$School of Fundamental Physics and Mathematical Sciences, Hangzhou Institute for Advanced Study, UCAS, Hangzhou 310024, China\\
}

\date{\today}

\begin{abstract}
We compute the scattering angle for a scalar neutral probe undergoing unbound motion around a Topological Star, including self-force effects.  Moreover we identify the `electro-magnetic' source of the background as Papapetrou Field compatible with the isometries and characterize Topological Stars by studying their sectional curvature, geometric transport along special curves and the gravitational energy content in terms of the super-energy tensors.

\end{abstract}

\maketitle
\section{Introduction}
\label{Intro}

As in any physical systems with local symmetry, observables in gravity are related to gauge-invariant quantities. For a standard two-body system (a pair of massive, spinning objects) in General Relativity (GR) one considers usually energy and angular momentum as fundamental gauge-invariant quantities, but there exist related ones like redshift, period and periastron advance  for bound motion  or the scattering angle for unbound motion.
Recently gravitational scattering has become particularly important because the scattering angle can be computed by using different approaches: Post-Newtonian (PN), summarized e.g. in Ref. \cite{Blanchet:2013haa},  and Post-Minkowskian (PM), started long ago with a number of seminal papers \cite{Westpfahl:1969ea,Westpfahl:1979gu,Westpfahl:1980mk,Bel:1981be} and recently reappraised (and popularized) by Ref. \cite{Damour:2016gwp},  approximations on the classical GR side, Effective-Field-Theory (EFT) see e.g. \cite{Foffa:2016rgu,Foffa:2019hrb,Porto:2016pyg,Kalin:2020mvi},  and finally Amplitudes (AMP) from the high energy physics side (see e.g., \cite{Bern:2019crd,Bern:2021dqo,Herrmann:2021tct,Bjerrum-Bohr:2021din,Bern:2021yeh,Dlapa:2021vgp,Dlapa:2022lmu,Damgaard:2023ttc,Dlapa:2024cje,Klemm:2024wtd,Driesse:2024xad}).
Actually,  the last years  have witnessed a positive synergy between the two approaches which have been also used to make predictions and also check the results of one approach against the other (see e.g., \cite{Bini:2021gat,Bini:2022enm}). 

Remarkably, waveforms and other observables can be also determined exploiting the relation between the wave equations governing gravitational perturbations and quantum Seiberg-Witten curves for $N=2$ supersymmetric gauge theories \cite{Aminov:2020yma,Bianchi:2021xpr,Bianchi:2021mft,Bonelli:2021uvf,
Bonelli:2022ten,Bianchi:2022qph,Consoli:2022eey, Bautista:2023sdf,DiRusso:2024hmd,Cipriani:2025ikx}

Here we study scattering processes in an electro-vacuum spacetime, namely in the case in which the source of the spacetime is not a neutral black hole (BH) but has an electromagnetic structure.  Our aim is that of identifying interesting signatures of the structure of the bodies involved in the scattering process.

We will mainly focus on the spacetime of a Top(ological) Star (TS) which -- for a specific choice of the parameters -- is a smooth horizonless solution of Einstein-Maxwell theory in $D=5$, briefly reviewed in Section \ref{TopSmetric} \cite{Bah:2020ogh,Heidmann:2023ojf,Bianchi:2023sfs,Bianchi:2024vmi,Bianchi:2024rod,Cipriani:2024ygw,Dima:2024cok,Bena:2024hoh,DiRusso:2025lip,Dima:2025zot}. We will identify the electro-magnetic source in terms of a Papapetrou Field \cite{Fayos:1999de}, compatible with the Killing structure (isometry) of the background, as discussed in Section \ref{KillPapa}, and characterize TS by studying their sectional curvature, geometric transport along special curves and gravitational energy content for a natural observer's choice, i.e. a static observer, in terms of the super-energy tensors, introduced by Bel and Robinson long ago 
\cite{bel58,bel59,bel62,Bel:1959uwe,rob97}
and reviewed in Section \ref{super-energy}.

We will then discuss the scattering process of a neutral, massive, spinless test particle  moving on an equatorial geodesic. We will provide explicit, analytic results for the scattering angle (a natural gauge-invariant quantity characterizing the process) by using a large angular momentum expansion limit. First, in Section \ref{TopSmetric}, we will consider simple geodetic motion that can be restricted to the equatorial plane thanks to spherical symmetry. Then, in Section \ref{scalSFeffectc}, relying on our recent study for unbound orbits in the background of a Top Star \cite{Bianchi:2024rod}, we will include scalar self-force effect that correct the geodetic result quite significantly. As a sanity check we will compare our results with the well-known ones for a Schwarzschild BH that represents a \lq limit' of the TS solution. We will compute both the conservative and the dissipative parts of the scattering angle. 

We will conclude in Section \ref{Conclusion} with comments on the inclusion of GW or electromagnetic emission in order to compute gravitational and electromagnetic self-force effects and massless scalar emission and draw some directions for future investigation in the case of Rotating Top Stars where most of the work is still in progress. 
Some cumbersome formulae can be found in the associated ancillary file.

\section{Killing vectors and Papapetrou Fields}
\label{KillPapa}

Consider a static and spherically symmetric metric of the type
\bea
\label{met_gen_d}
ds^2&=&-f_t(r)dt^2+f_r(r)dr^2 +f_{\Omega}(r)(d\theta^2+\sin^2\theta d\phi^2)\nonumber\\
&+& (D-4)f_y(r)dy^2\,,
\eea 
with coordinates $(t,r,\theta,\phi,y)$ adapted to the Killing symmetries of the spacetime, that may be viewed as a non-vacuum solution of Einstein equations with sources.
Here $D$ denotes the spacetime dimension: 4 in GR but 5 in the case of a Top(ological) Star (TS) spacetime, that will be the main focus of our analysis starting from Section \ref{TopSmetric}.
Sometimes, to ease comparisons of the TS spacetime with other 4D spacetimes we will consider a $D=4$ TS metric, i.e., a 4D spacetime with metric  the one induced on the leaves $y=$constant.

Noticeably, when the background admits as a source an electromagnetic field, the latter is simply related to the Killing structure of the spacetime, in the sense that
the vector potential $A$ has components (eventually separated in their dependence on $r$ and $\theta$) only along the Killing directions\footnote{In principle one can include a component $A_y(r,\theta)$, but we will refrain from doing so.}
\beq
A=A_t(r,\theta) dt +A_\phi(r,\theta) d\phi\,. 
\eeq
Let us recall that the Killing vectors themselves play a role in this context. In fact, if $\xi$ is a Killing vector field then by definition (since $\xi_{(\alpha;\beta)}=0$ and hence $\xi_{\alpha;\beta}=F_{\alpha\beta}$ is antisymmetric, and  satisfies Maxwell's equations with sources,
\beq
F_\xi^{\alpha\beta}{}_{;\beta}=R^\alpha{}_\beta \xi^\beta=J^\alpha\,.
\eeq
$F_\xi^{\alpha\beta}$ is termed a Papapetrou Field (PF) \cite{Fayos:1999de}, and in vacuum ($R_{\alpha\beta}=0$) they are such that $F_\xi^{\alpha\beta}{}_{;\beta}=0$.
PF have been used to analyze exact solutions of Einstein's field equations, and in the study of black holes in the presence of
external electromagnetic fields \cite{Wald:1974np}. In particular, it turns out that the Kerr-Newman
electromagnetic field is a PF generated by the time-like Killing vector field of the Kerr
metric.
Moreover, PFs provide us with a link between the Killing
symmetries and the algebraic structure of the spacetime, which is a subject treated
only occasionally in the literature \cite{Fayos:1999de,Fayos:2002pa}. 

This relationship can be found
by studying the alignment of the principal directions of the PF with those
of the gravitational field, i.e., of the Riemann tensor (which in vacuum reduces to the
Weyl tensor). In this sense, a remarkable example is the type D Kerr metric: the principal directions of the PF associated with the
time-like Killing vector field coincide with the two repeated principal directions of the spacetime. This link between Killing symmetries and algebraic structure represents a
powerful tool for the study (and the search) of vacuum gravitational fields. 
Combining two such fields with coefficient functions one may find conditions such that source-free Maxwell's equations be satisfied. 

\section{Super-energy tensors}
\label{super-energy}
The main interest of our present investigation is the spacetime of a TS which
we can also characterize  by studying their gravitational energy content in terms of super-energy tensors, as those due to Bel and Bel-Robinson long ago \cite{bel58,bel59,bel62,Bel:1959uwe,rob97,Deser:1977zz,Breton:1993du,Mashhoon:1996wa,Bonilla:1997ink,Bergqvist:1998b,Mashhoon:1998tt,Senovilla:1999xz,bini-jan-miniutti}
.

The Bel and Bel-Robinson tensors are built with the Riemann tensor and the Weyl tensor, respectively, and are both divergence-free in vacuum spacetimes, where they coincide. 
Their mathematical properties are reviewed, e.g., in Refs. \cite{Bonilla:1997ink,Senovilla:1999xz}.
We will consider the following gravitational (Bel) super-energy tensor (see Eq. 15.29 of Ref. \cite{Misner:1973prb})
\beq
\label{BelR}
 T^{\rm (g)}{}_{\alpha\beta} {}^{\gamma\delta}
 =
    R_{\alpha\rho\beta\sigma} R^{\gamma\rho\delta\sigma}
   + {}^* R_{\alpha\rho\beta\sigma} {}^* R^{\gamma\rho\delta\sigma} 
\ ,
\eeq
defined in terms of the Riemann tensor $R_{\alpha\rho\gamma\delta}$ and its left dual 
\beq
\label{LeftDual}
 {}^* R_{\alpha\rho\beta\sigma} =\frac12 {\varepsilon}_{\alpha\rho\gamma\delta}R^{\gamma\delta}{}_{\beta\sigma} 
\ ,
\eeq
which is more natural in the case of nonempty spacetimes. In an empty spacetime, the latter tensor is defined exactly in the same way by using the Weyl tensor and enjoys the following properties
\begin{enumerate}
\item It is symmetric and traceless in all pair of   indices; 
\item It is divergence-free $ \nabla_{\alpha}T^{\rm (g)}{}^{\alpha} {}_{\beta\gamma\delta}=0$.
\end{enumerate}

Indeed, the analogous quantity (Bel-Robinson) defined in terms of the  Weyl tensor $C_{\alpha\beta\gamma\delta}$ and its dual carries a similar information, but without any reference to the spacetime sources. We will not insist on studying explicitly the differences between these two tensors (already discussed in the literature \cite{Bonilla:1997ink,Senovilla:1999xz}), 
and take the Bel tensor as bearer of  (qualitative, more than quantitative) information concerning the observer-dependent gravitational energy and momentum content of a given spacetime (and a given observer).  

As it is well known, in analogy with electromagnetism, by using the Bel tensor one can introduce the  super-energy density and super-momentum density (or super-Poynting, spatial) vector \cite{bel58,bel59,bel62,Bel:1959uwe} with respect to a given observer with 4-velocity $u$ 
\begin{eqnarray}
\label{Belsuperendef}
 \mathcal{E}^{\rm{(g)}}(u)
 &=&  T^{\rm (g)}_{\alpha\beta\gamma\delta} u^\alpha u^\beta u^\gamma u^\delta\ ,
\nonumber\\
 \mathcal{P}^{\rm{(g)}}(u)_\alpha
 &=& P(u)^\epsilon{}_\alpha T^{\rm (g)}_{\epsilon\beta\gamma\delta} u^\beta u^\gamma u^\delta
\,.
\end{eqnarray}
with $u$ unitary, timelike and future-pointing and $P(u)=g+u\otimes u$ the projection operator orthogonally to $u$.

It is convenient to introduce the notation
\beq
X_{\alpha\beta\ldots}u^\alpha=X_{{\mathbf u}\beta\ldots}
\eeq
for indices contracted with $u$ and to recall the notion of electric and magnetic parts of the Riemann tensor
 \bea
\label{EHriemann}
{\tt E}(u)_{\alpha\beta}&=&R_{\alpha\mu\beta\nu}u^\mu u^\nu=R_{\alpha {\bf u}\beta {\bf u}}\,,\nonumber\\
{\tt H}(u)_{\alpha\beta}&=& {}^*R_{\alpha\mu\beta\nu}u^\mu u^\nu={}^*R_{\alpha {\bf u}\beta {\bf u}}\,.
 \eea
We have then
\begin{eqnarray}
\label{Belsuperendef2}
 \mathcal{E}^{\rm{(g)}}(u)
 &=&  R_{{\bf u} \rho {\bf u} \sigma} R^{{\bf u} \rho {\bf u} \sigma}
   + {}^* R_{{\bf u} \rho {\bf u} \sigma} {}^* R^{{\bf u} \rho {\bf u} \sigma} \ ,
\nonumber\\
&=& {\tt E}(u)_{\rho\sigma}{\tt E}(u)^{\rho\sigma}  
   + {\tt H}(u)_{\rho\sigma}{\tt H}(u)^{\rho\sigma} \ ,
\nonumber\\
 \mathcal{P}^{\rm{(g)}}(u)_\alpha
 &=& P(u)^\epsilon{}_\alpha (R_{\epsilon\rho {\bf u} \sigma} R^{{\bf u} \rho {\bf u} \sigma}
   + {}^* R_{\epsilon\rho {\bf u} \sigma} {}^* R^{{\bf u} \rho {\bf u} \sigma})  \nonumber\\
&=& P(u)^\epsilon{}_\alpha (R_{\epsilon\rho {\bf u} \sigma} {\tt E}(u)^{\rho\sigma}
   + {}^* R_{\epsilon\rho {\bf u} \sigma} {\tt H}(u)^{\rho\sigma} )  
\,.\nonumber\\
\end{eqnarray}
Here ${\tt E}(u)_{\rho\sigma}$  is spatial with respect to $u$ (any contraction with $u$ vanishes), symmetric (but not tracefree),  while ${\tt H}(u)_{\rho\sigma}$ is spatial with respect to $u$ and not symmetric (but tracefree).
A useful method to work with tensors in index-free notation when one needs to split their symmetric and antisymmetric parts is the following
\bea
{\tt H}_{\rm S}(u)&\equiv& [{\tt H}(u)^{(\alpha\beta)}]\,,\nonumber\\
{\tt H}_{\rm A}(u)&\equiv& [{\tt H}(u)^{[\alpha\beta]}]\,.
\eea
Therefore
\begin{eqnarray}
\label{Belsuperendef3}
 \mathcal{E}^{\rm{(g)}}(u)
&=& {\rm Tr}\, [ {\tt E}(u)\cdot {\tt E}(u)]\nonumber\\  
 &  +& {\tt H}(u)_{\rho\sigma}[{\tt H}(u)^{(\rho\sigma)}+{\tt H}(u)^{[\rho\sigma]}] \ ,
\nonumber\\
&=& {\rm Tr}\, [ {\tt E}(u)\cdot {\tt E}(u)+{\rm Tr}\, [ {\tt H}_{\rm S}(u)\cdot {\tt H}_{\rm S}(u)]\nonumber\\  
&-&{\rm Tr}\, [{\tt H}_{\rm A}(u)\cdot  {\tt H}_{\rm A}(u)] \ ,
\end{eqnarray}
where matrix multiplication is meant as a contraction of the rightmost lower index of the first tensor with leftmost upper index of the second, 
\beq
[A\cdot B]_\alpha{}^\gamma=A_{\alpha\beta}B^{\beta\gamma}\,,\qquad {\rm Tr}[A\cdot B]=[A\cdot B]_\alpha{}^\alpha\,.
\eeq
Let us now recall that a spatial antisymmetric 2-tensor can be fully represented by a vector, labeled by a $V$ (exactly as the magnetic field 2-tensor is equivalent to a corresponding vector), i.e., 
\beq
[{\tt H}_{\rm A}(u)]_{\alpha\beta}={\varepsilon}_{\bf u}{}_{\alpha\beta\gamma}[{\tt H}_{\rm V}(u)]^\gamma\,,
\eeq
with ${\varepsilon}_{\bf u}{}_{\alpha\beta\gamma} 
= u^\delta \,{\varepsilon}_{\delta\alpha\beta\gamma}$ the unit volume 3-form in the local rest space (LRS) of the observer $u$.
With the above 3-form one can also define a spatial cross product for two symmetric spatial tensor fields ($A, B$) 
\beq
[A\times_u B]_{\alpha} = {\varepsilon}_{\bf u}{}_{\alpha\beta\gamma} A^{\beta} {}_{\delta}\, B^{\delta\gamma}\,.
\eeq
We can further manipulate the term
\bea
{\rm Tr}\, [{\tt H}_{\rm A}(u)\cdot  {\tt H}_{\rm A}(u)]&=&{\varepsilon}_{\bf u}{}_{\alpha\beta\gamma}[{\tt H}_{\rm V}(u)]^\gamma {\varepsilon}_{\bf u}{}^{\beta\alpha\rho}[{\tt H}_{\rm V}(u)]_\rho \nonumber\\
&=& -2\delta_\gamma^\rho [{\tt H}_{\rm V}(u)]^\gamma [{\tt H}_{\rm V}(u)]_\rho\nonumber\\
&=& -2 |{\tt H}_{\rm V}(u)|^2\,.
\eea
Finally, one writes
\begin{eqnarray}
\label{Belsuperendef4}
 \mathcal{E}^{\rm{(g)}}(u)
&=& {\rm Tr}\, [ {\tt E}(u)\cdot {\tt E}(u)]+{\rm Tr}\, [ {\tt H}_{\rm S}(u)\cdot {\tt H}_{\rm S}(u)]\nonumber\\  
&+&2  |{\tt H}_{\rm V}(u)|^2 \,,
\eea
which is the gravito-electromagnetic counterpart ($1+3$ split version) to the familiar electromagnetic field energy
\beq
 \mathcal{E}^{\rm{(em)}}(u)
=|{\tt E}(u)|^2+|{\tt H}(u)|^2\,.
\eeq
Proceeding along the same lines one can compute the $1+3$ split version of the super-momentum. However, as discussed below, its spatial part vanishes identically and we will not insist on formal analogies any further and pass to consider TSs.

\section{Top Star metric} 
\label{TopSmetric}

A Topological Star (TS) is described by the following $D=5$ metric
\bea\label{metric}
ds^2&=&-f_s(r)dt^2+\frac{dr^2}{f_s(r)f_b(r)}\nn\\
&+& r^2(d\theta^2+\sin^2\theta d\phi^2)+f_b(r)dy^2\,,
\eea
with 
\be
f_{s,b}(r)=1-\frac{r_{s,b}}{r}.
\ee
The coordinate $y$ is compact $y\sim y+2\pi R_y$. The metric \eqref{metric} is a magnetically charged solution of $D=5$ Einstein-Maxwell system of equations, where the electromagnetic field is given by
\beq
A=-P\cos\theta d\phi\,,\qquad F=P\sin\theta d\theta\wedge d\phi\,,
\eeq
and with 
\beq
P^2=\frac{3r_b r_s}{2\kappa_5^2}\,,
\eeq
 representing the magnetic charge.
Moreover, $F^2=-F_{\alpha\beta}F^{\alpha\beta}=\frac{2P^2}{r^4}$ and
\beq
{}^*F=\frac{2P}{r^2} dt\wedge dr\wedge dy\,,
\eeq
and
\beq
T_{\mu\nu}=-\frac{P^2}{2r^4}   g_{\mu\nu}+\frac{P^2}{r^2}[\delta_\mu^\theta \delta_\nu^\theta+\sin^2\theta \delta_\mu^\phi \delta_\nu^\phi]\,.
\eeq
Actually, with this form of the metric there exist two different regimes: 1) black string (BS): $r_b<r_s$, with an event horizon at $r=r_s$; 2) TS: $r_s<r_b$, a smooth
horizonless solution if   $y$ is periodically identified with period $2\pi R_y$ ($R_y=2r_b^{3/2}/(r_b-r_s)^{1/2}$). Furthermore, stability against (metric) linear perturbations requires $r_b<r_s<2r_b$
in the BS case and $r_s<r_b<2r_s$ in the TS one \cite{Bah:2020ogh,Heidmann:2023ojf,Bianchi:2023sfs}.

After dimensional reduction to $D= 4$, the solution shows a naked singularity
and has a mass
\beq
G_4M_{\rm TS} = \frac{r_s}2  + \frac{r_b}4 \,, 
\eeq
with
\beq
8\pi G_4=\kappa_4^2=\frac{\kappa_5^2}{2\pi R_y}=\frac{8\pi G_5}{2\pi R_y}\,.
\eeq
Hereafter we will set $G_4\equiv G = 1 = c$. For $r_b = 0$, and
thus $r_s = 2GM_{\rm TS}$, the resulting singular solution is 
Schwarzschild BH$\times S_1$.
Often we will find it convenient to denote
$r_s=2M$, $r_b=\alpha r_s=2\alpha M$,
with $M$ a common length scale, not to be confuse with $M_{\rm TS}$, the mass of the TS, unless for $r_b=0$. In fact,
\beq
G_4M_{\rm TS} =\frac{r_s}{4}(\alpha+2)=\frac{M}{2}(\alpha+2) \,. 
\eeq

\subsection{Sectional curvatures}

Let us consider the following sections of the TS spacetime:

\begin{enumerate}
  \item $t-r$ section, with (Lorentzian) 2-metric
\beq
{}^{(2)}ds^2 =-f_s(r)dt^2+\frac{dr^2}{f_s(r)f_b(r)}\,.
\eeq
The associated Kretschmann invariant is
\beq
{}^{(2)}K_{t-r}=R_{\alpha\beta\gamma\delta} R^{\alpha\beta\gamma\delta} =\frac{r_s^2(4 r-5 r_b)^2}{4 r^8}\,,
\eeq 
showing a geometrical role for the radius $r=\frac{5}{4}r_b$.

 \item $r-\phi$ section, with (Euclidean) 2-metric
\beq
{}^{(2)}ds^2 = \frac{dr^2}{f_s(r)f_b(r)}+r^2d\phi^2\,.
\eeq
The associated Kretschmann invariant is
\beq
{}^{(2)}K_{r-\phi}=\frac{[(r_s+r_b) r-2 r_s r_b]^2}{ r^8}\,,
\eeq 
showing a geometrical role for the radius $r=\frac{2r_sr_b}{(r_s+r_b)}=\frac{2}{\frac{1}{r_s}+\frac{1}{r_b}}$.

 \item $r-y$ section, with (Euclidean) 2-metric
\beq
{}^{(2)}ds^2 = \frac{dr^2}{f_s(r)f_b(r)}+f_b(r)dy^2\,.
\eeq
The associated Kretschmann invariant is
\beq
{}^{(2)}K_{r-y}=\frac{r_b^2(4 r-5 r_s)^2}{4 r^8} \,,
\eeq 
showing a geometrical role for the radius $r=\frac{5}{4}r_s$, namely $r_s\leftrightarrow r_b$ exchange their roles when passing from ${}^{(2)}K_{t-r}$ to ${}^{(2)}K_{r-y}$, as expected.

\end{enumerate}
\subsection{Static observers, split of the Riemann and super-energy  tensors}

 Let us consider the constant $y$ section of the TS spacetime. With the Riemann tensor induced on this hypersurface we can form the super-energy as measured by the static observer,
\beq
\label{stat_obs}
u=\frac{1}{\sqrt{-g_{tt}}}\partial_t\,.
\eeq
We find
\bea
\mathcal{E}^{\rm{(g)\, TS}}(u)&=& \frac{(24 r^2 - 56 r r_b + 33 r_b^2) r_s^2}{16 r^8}\nonumber\\
&=& \frac{3 r_s^2}{2 r^6} - \frac{7 \alpha r_s^3}{2 r^7} + \frac{33 \alpha^2 r_s^4}{16 r^8}\,,
\eea
with $r_b=\alpha r_s$, as above.

\begin{figure}
\includegraphics[scale=0.8]{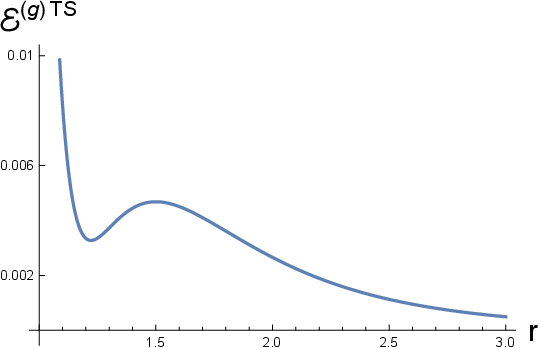} 
\caption{\label{super-energy_TS} super-energy as measured by the static observer and stored on the leaves $y=$ constant for a TS spacetime plotted as function of the radial coordinate $r$ for
$r_b=1$ and $r_s=0.8$. The extrema are located at
$r_{\rm max}=\frac{3r_b}{2}$ (corresponding to $\mathcal{E}^{\rm{(g)\, TS}}(r_{\rm max})=\frac{16}{ 2187\alpha^6 r_s^4}\approx 0.004682$) and  $r_{\rm min}=\frac{11 r_b}{9}$ (corresponding to $\mathcal{E}^{\rm{(g)\, TS}}(r_{\rm min})=\frac{1594323}{311794736\alpha^6 r_s^4}\approx 0.003273$).}
\end{figure}

It dimensionless version $\hat {\mathcal E}^{\rm{(g)\, TS}}=r_s^4 {\mathcal E}^{\rm{(g)\, TS}}$ is then conveniently written as
\bea
\hat {\mathcal E}^{\rm{(g)\, TS}} &= &\frac{3}{2}x^6 - \frac{7 \alpha }{2}x^7 + \frac{33 \alpha^2 }{16} x^8\nonumber\\
&=& \frac{3}{2}x^6 \left(1 - \frac{7 }{3} \alpha x  + \frac{11  }{8}\alpha^2 x^2\right)\nonumber\\
&\equiv & \hat {\mathcal E}^{\rm{(g)\, Schw}}\, A(\alpha x)\,,
\eea 
where we have defined $x=\frac{r_s}{r}$. With respect to the case $\alpha=0$ (Schwarzschild on the leaves $y$ = constant) we have here the amplifying factor 
\beq
A(\alpha x)=1 - \frac{7 }{3} \alpha x  + \frac{11  }{8}\alpha^2 x^2\,,
\eeq 
continuously increasingly with a parabolic behavior starting for the value
$\frac{1}{24}=0.041\bar 6$, obtained for $\alpha x=1$, corresponding to $r=r_b$.

\subsection{Geometric transport along special orbits}

Special insight into the gravitational field of the TS is obtained by studying  particles' motion. In this subsection we will consider timelike, equatorial circular orbits in the $t-\phi$ 2-plane at fixed radius $r=r_0>r_b$, not necessarily geodesics, i.e. characterized by the four velocity
\beq
U_{\rm circ}=\Gamma (\partial_t +\Omega \partial_\phi )\,,
\eeq
with
\beq
\Gamma=\left[1-\frac{r_s}{r_0}-\Omega^2 r_0^2\right]^{-1/2}\,,
\eeq
and parametric equations 
\beq
t=\Gamma \tau\,,\quad r=r_0\,,\quad \theta=\frac{\pi}{2}\,,\quad \phi=\Gamma \Omega \tau\,,\quad y=y_0\,.
\eeq

Here $\Omega$ is a free parameter which defines the family of these orbits, existing if
\beq
-\Omega_* \le \Omega \le \Omega_*\,,
\eeq
with
\beq
\Omega_*=\frac{1}{r_0}\left(1-\frac{r_s}{r_0}\right)^{1/2}\,.
\eeq
For instance, when $\Omega=\Omega_{\rm g}=\sqrt{r_s/(2r_0^3)}$ the orbits are geodesics.

Let us introduce an orthonormal frame adapted to the static observers \eqref{stat_obs}
\bea
\label{nat_frame}
e_{\hat t}&=&u=\frac{1}{\sqrt{-g_{tt}}}\partial_t \,,\quad e_{\hat r}=\frac{1}{\sqrt{g_{rr}}}\partial_r  \,,\quad e_{\hat \theta}=\frac{1}{\sqrt{g_{\theta\theta}}}\partial_\theta  \,,\nonumber\\
e_{\hat \phi}&=&\frac{1}{\sqrt{g_{\phi\phi}}}\partial_\phi  \,,\quad e_{\hat y}=\frac{1}{\sqrt{g_{yy}}}\partial_y\,.   
\eea
Instead of the $\Omega$ parametrization it can be convenient to have a velocity parametrization writing
\beq
U_{\rm circ}=\gamma (e_{\hat t} +v e_{\hat \phi})\,,
\eeq
such that
\beq
\Gamma=\frac{\gamma}{\sqrt{-g_{tt}}}\,,\quad v=\frac{\sqrt{g_{\phi\phi}}}{\sqrt{-g_{tt}}}\,,
\eeq
where now the physical velocity (signed magnitude of the relative velocity of $U_{\rm circ}$ with respect to the static observers) varies as $v\in [-1,1]$.
Finally, the acceleration of such orbits is purely radial
\bea
a(U_{\rm circ})&=&\nabla_{U_{\rm circ}}U_{\rm circ}\nonumber\\
&=& \frac{\Gamma^2 f_s(r_0)f_b(r_0)}{r_0^2}(M-\Omega^2 r_0^3)\partial_r\,,
\eea
and changes its sign at $\Omega=\Omega_{\rm g}$ (i.e., it is directed radially inward only at those radii and those frequencies such that $\Omega>\Omega_{\rm g}$).

\subsubsection{Parallel transport}

Consider a generic vector $X=X^\alpha \partial_\alpha$ and study the equation for parallel transport of $X$ along $U_{\rm circ}$,
\beq
\frac{dX^\alpha}{d\tau}+\Gamma^\alpha{}_{\mu\nu}X^\mu U_{\rm circ}^\nu=0\,.
\eeq
the transport is trivial for the $\theta$ and $\phi$ components of $X$ and reduces to $X^\theta(\tau)=X^\theta(0)$ and $X^y(\tau)=X^y(0)$. We can then assume $X^\theta(0)=X^y(0)=0$. The other components are such that
\bea
\label{sys_paral}
\frac{dX^{\hat t}}{d\tau}&=& -\frac{r_s}{2r_0^2}\Gamma \sqrt{f_b(r_0)} X^{\hat r}\,,\nonumber\\
\frac{dX^{\hat r}}{d\tau}&=& \Gamma \Omega \sqrt{f_s(r_0)f_b(r_0)} X^{\hat \phi}-\frac{r_s }{2 r_0^2}\Gamma\sqrt{f_b(r_0)} X^{\hat t}\,,\nonumber\\
\frac{dX^{\hat \phi}}{d\tau}&=& -\Gamma \Omega \sqrt{f_s(r_0)f_b(r_0)} X^{\hat r}\,,
\eea
where the hat-components are taken with respect to the natural frame \eqref{nat_frame}.
In fact, along the above equatorial circular orbit
\bea
X^t(\tau) &=& \frac{X^{\hat t}(\tau)}{\sqrt{f_s(r_0)}}\,,\qquad X^{\phi}(\tau) =\frac{X^{\hat\phi}(\tau)}{r_0}\,, \nonumber\\
X^r(\tau) &=& X^{\hat r}(\tau)\sqrt{f_s(r_0)f_b(r_0)}\,, \nonumber\\
X^{\theta}(\tau) &=& \frac{1}{r_0}X^{\hat \theta}(\tau)\,, \qquad
X^y(\tau) =\frac{X^{\hat y}(\tau)}{\sqrt{f_b(r_0)}}\,.
\eea
The system \eqref{sys_paral} can be conveniently written as
\bea
\label{sys_paral_formal}
\frac{dX^{\hat t}}{d\tau}&=& -{\mathcal A}  X^{\hat r}\,,\nonumber\\
\frac{dX^{\hat r}}{d\tau}&=& {\mathcal B} X^{\hat \phi}-{\mathcal A}  X^{\hat t}\,,\nonumber\\
\frac{dX^{\hat \phi}}{d\tau}&=& -{\mathcal B} X^{\hat r}\,,
\eea
with
\bea
{\mathcal A}&=& \frac{r_s}{2r_0^2}\Gamma \sqrt{f_b(r_0)}\,,\nonumber\\
{\mathcal B}&=& \Gamma \Omega \sqrt{f_s(r_0)f_b(r_0)}\,,
\eea
and 
\beq
\zeta={\mathcal A}^2-{\mathcal B}^2=\frac{\Gamma^2 f_b(r_0) (-r_0^4 \Omega^2 f_s(r_0) +\frac{r_s^2}{4})}{ r_0^4 }\,.
\eeq

These equations decouple simply by differentiating the radial one with respect to $\tau$ and using in it the other two equations, obtaining
\beq
\frac{d^2 X^{\hat r}}{d\tau^2}=\zeta  X^{\hat r}\,.
\eeq

As in the case of black holes \cite{Bini:2004sy}, the solutions are of the type
\begin{enumerate}
  \item A spatial rotation from the initial data for all $\Omega$ such that $\zeta<0$;
  \item A boost for all $\Omega$ such that $\zeta>0$;
  \item A null rotation for $\Omega=\pm \frac{r_s}{2r_0^2 \sqrt{f_s(r_0)}} $ , corresponding to $\zeta=0$.
\end{enumerate}
For example, in the case $\zeta<0$ we find
\beq
\frac{d^2 X^{\hat r}}{d\tau^2}=-|\zeta | X^{\hat r}\,,
\eeq
which implies
\beq
X^{\hat r}(\tau)=\frac{X^{\hat r}{}'(0)}{\sqrt{|\zeta|}} \sin(\sqrt{|\zeta|}\tau )+X^{\hat r}(0) \cos(\sqrt{|\zeta|}\tau )\,,
\eeq
a periodic function with (minimum) period $\frac{2\pi}{\sqrt{|\zeta|}}$. A number of $k$ full orbital revolutions corresponds  to
\beq
2\pi k =\Gamma \Omega \tau_*\,,\quad k \in {\mathbb Z}\,,
\eeq
and therefore
\beq
X^{\hat r}(\tau_*)=\frac{X^{\hat r}{}'(0)}{\sqrt{|\zeta|}} \sin(\sqrt{|\zeta|}\frac{2k\pi}{\Gamma \Omega} )+X^{\hat r}(0) \cos(\sqrt{|\zeta|}\frac{2k\pi}{\Gamma \Omega} )\,.
\eeq
A relation for \lq\lq holonomy invariance" can then be written assuming
\beq
\label{hol_inv}
\sqrt{|\zeta|}\frac{2\pi k }{\Gamma \Omega}=2n\pi\,,\qquad n \in {\mathbb Z}\,,
\eeq
ensuring $X^{\hat r}(\tau_*)=X^{\hat r}(0)$.
Eq. \eqref{hol_inv} can be also written as
\beq
\label{hol_inv2}
\frac{\sqrt{|\zeta|}}{\Gamma \Omega}= \frac{n}{k}\,, 
\eeq
defining special frequencies/radii where this condition is satisfied.  [Note that a negative $\Omega$ is allowed and corresponds to a circular orbit rotating clockwise.] A  numerical study  of Eq. \eqref{hol_inv2} is shown in Fig. \ref{fig:holon}.

\begin{figure}
\includegraphics[scale=0.35]{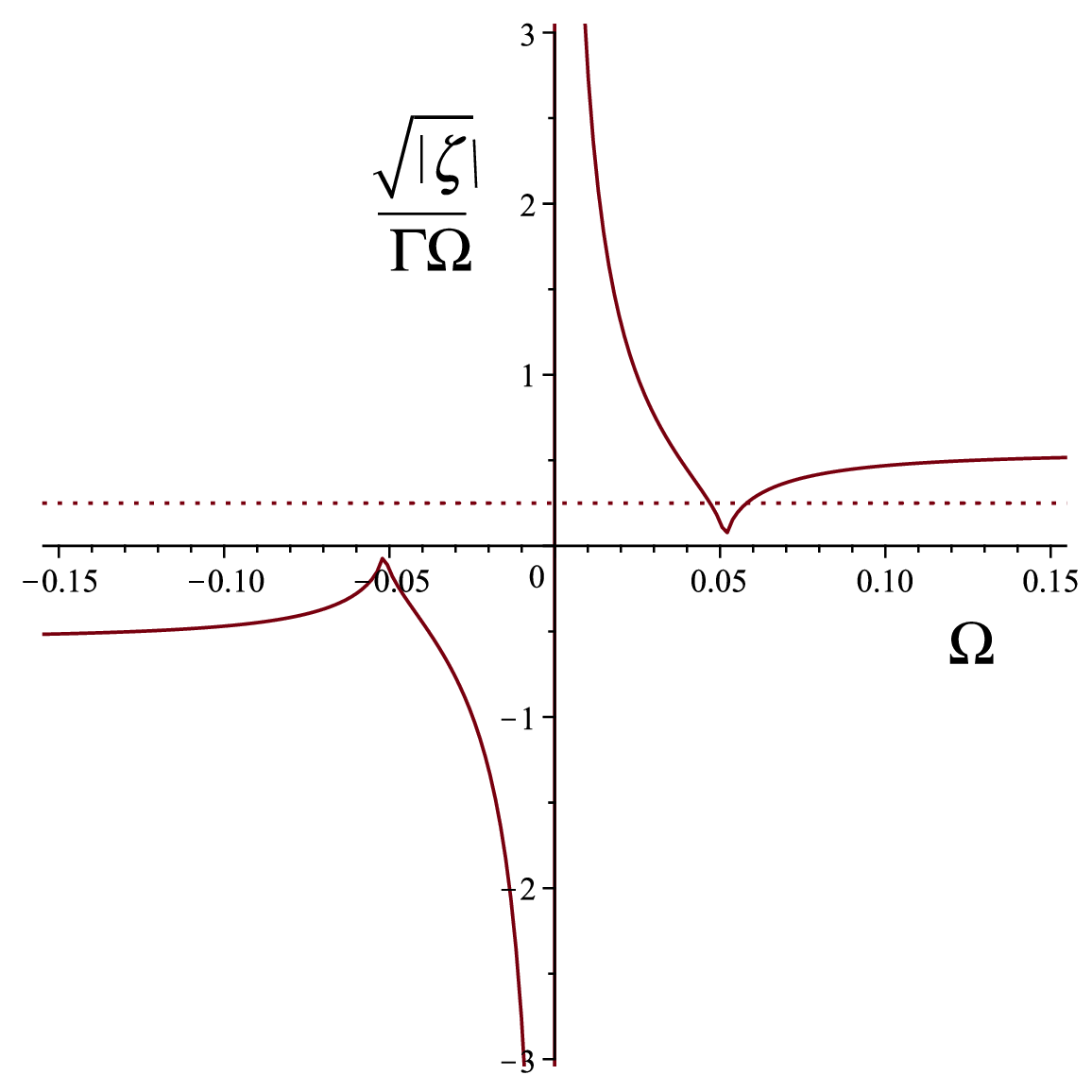} 
\caption{\label{fig:holon} The LHS of Eq. \eqref{hol_inv2} is plotted as a function of $\Omega\in [-\Omega_*,\Omega_*]$ for  $\alpha=5/4$, $r_0=5M$, $M=1$. The numerical value of $\Omega_*$ in this case is $0.1549$. The dotted line superposed corresponds to the right-hand-side of Eq. \eqref{hol_inv2} with $n=1$, $k=4$.}
\end{figure}

\subsubsection{Fermi-Walker transport}

Let us assume that the vector $X$ defined above (ignoring $\theta$ and $y$ components) is also orthogonal to $U_{\rm circ}$, namely
\beq
X^t(\tau) = \frac{\Omega r_0^2}{f_s(r_0)} X^\phi(\tau)\,,
\eeq
and let us study the Fermi-Walker transport law of $X$ along $U_{\rm circ}$
\beq
P(U_{\rm circ})_\alpha{}^\beta \nabla_{U_{\rm circ}} X_\beta=0\,.
\eeq
The evolution equations read
\bea
\frac{dX^{\hat \phi}}{d\tau}&=&  \Gamma^3\Omega \sqrt{f_s(r_0)f_b(r_0)}\left(1-\frac{3M}{r_0}\right)
  X^{\hat r}\,,\nonumber\\
\frac{dX^{\hat r}}{d\tau}&=& \Gamma   \frac{\sqrt{f_b(r_0)}}{\sqrt{f_s(r_0)}}\left(1-\frac{3M}{r_0}\right) X^{\hat \phi} \,,
\eea
leading to
\beq
\frac{d^2 X^{\hat r}}{d\tau^2}+\Omega_{\rm fw}^2  X^{\hat r}=0\,,
\eeq
where
\beq
\Omega_{\rm fw}=   \Gamma^2 \Omega \sqrt{f_b(r_0)}\left|1-\frac{3M}{r_0}\right| \,.
\eeq
The solution for $X^{\hat r}$ in this case corresponds in general to a rotation  or a null rotation  (at $r_0=3M$) (but never a boost). For example, if $r_0\not =3M$
\bea
X^{\hat r}(\tau)&=&\frac{X^{\hat r}{}'(0)}{\Omega_{\rm fw}}\sin\left(\Omega_{\rm fw}\tau\right)+X^{\hat r}(\tau) \cos\left(\Omega_{\rm fw} \tau\right)\,,\qquad
\eea
and one can study again the holonomy invariance properties.

\subsection{Hyperbolic-like geodesic motions and the scattering angle}

The geodesics of  the metric \eqref{metric} can be derived from the Lagrangian density
\bea
\mathcal{L}&=&\frac12 g_{\alpha\beta}\dot x^\alpha \dot x^\beta\,,
\eea
where a dot denotes differentiation with respect to an affine parameter $\lambda$ related to the proper time by $\tau=\lambda m $ with $m$ the mass of the particle (probe).
The corresponding Hamiltonian reads
\be\label{hamiltonian}
\mathcal{H}=\frac12 g^{\mu\nu}P_\mu P_\nu,\quad P_\mu=\frac{\partial \mathcal{L}}{\partial \dot{x}^\mu}\,,
\ee
with mass-shell condition $\mathcal{H}=-m^2/2$.

The conserved momenta are denoted as
\be
    P_t=-E,\quad P_y=p,\quad P_\phi=J\,.
\ee
The geodesic equations can be separated by introducing the Carter-like constant $K$, so that
\bea
    P_r^2  
&=& \frac{E^2 r^3}{(r-r_b)(r-r_s)^2}-\frac{K^2}{(r-r_b)(r-r_s)}\nn\\
    &-&\frac{r^2(p^2r+(r-r_b)m^2)}{(r-r_b)^2(r-r_s)}\,, \nn\\
    P_\theta^2  
&=& K^2-\frac{J^2}{\sin^2\theta}\,.
\eea
Due to the spherical symmetry of the background, we can focus on the equatorial motion only without loss of generality, $\theta=\frac{\pi}{2}$ ($P_\theta=0$) so that $K=J$. Furthermore, we will limit our considerations to the case $p=0$, implying
\bea
    p_r^2&\equiv& \frac{P_r^2}{m^2} 
=\frac{ \hat E^2 \hat r^3 - (j^2 +\hat r^2) (\hat r-2)}{(\hat r-2\alpha)(\hat r-2)^2} \,.
\eea  
where we have defined the dimensionless quantities
\beq
\hat E=\frac{E}{m c^2}\,,\qquad j=\frac{c J}{G M m}\,,\qquad \hat r =\frac{c^2 r}{GM}, 
\eeq
(mostly used with $c=1$)
besides 
\beq
\frac{r_s}{M}=2\,,\qquad \frac{r_b}{r_s}=\alpha\,.
\eeq
We will also write $\hat E=\gamma$ following standardly used notation.
Finally, let us pass from $\hat r$ to ${w}=\frac{1}{\hat r}$, so that
\bea
\label{prdef}
    p_r^2 (1-2{w})^2
&=& \frac{ \hat E^2   - (j^2{w}^2 +1) (1-2{w})}{(1-2\alpha {w})} \,.
\eea 
From Eqs. \eqref{hamiltonian}, we obtain
\bea
\frac{dr}{d\tau}&=&\frac{p_r(r-r_b)(r-r_s)}{r^2},\nonumber\\
\frac{d\phi}{d\tau}&=&\frac{M j}{r^2}=\frac{j}{M \hat r^2}\,,
\eea
that is
\beq
\frac{dr}{d\tau}=  p_r(1-2\alpha{w})(1-2{w})\,,\qquad \frac{d\phi}{d\tau}=\frac{j}{M }{w}^2\,.
\eeq
Consequently
\beq
\frac{d\phi}{d\hat r}=\frac{{w}^2 j}{p_r(1-2\alpha {w})(1-2{w})}\,,
\eeq
and
\beq
\frac{d\phi}{d{w}}=-\frac{j}{p_r(1-2\alpha {w})(1-2{w})}\,.
\eeq
Using Eq. \eqref{prdef} we find
\beq
\label{eq_scat_u}
\frac{d\phi}{d{w}}=-\frac{j}{ \sqrt{1-2\alpha {w}} \sqrt{ \hat E^2   - (j^2{w}^2 +1) (1-2{w}) }}\,,
\eeq
equivalent to
\beq
\frac{d\phi}{dr}=\frac{\partial p_r}{\partial j} \,.
\eeq
The scattering angle follows from integration along the full orbit,
\beq
\frac{\chi+\pi}{2}=\int_0^{{w}_{\rm max}} \frac{j}{ \sqrt{1-2\alpha {w}} }\frac{d{w}}{\sqrt{ \hat E^2   - (j^2{w}^2 +1) (1-2{w}) }}\,.
\eeq
where ${w}_{\rm max}={w}_{\rm max}(\hat E,j)$ is the largest solution of 
\beq
\label{wmax}
\hat E^2   - (j^2{w}^2 +1) (1-2{w}) =0 \, .
\eeq

We can proceed now evaluating Eq. \eqref{eq_scat_u}  in the large $j$ expansion limit. 
Let us rewrite conveniently the argument of the square root denoting $\hat E^2-1=2\bar E$
\bea
{\mathcal D}&=& \hat E^2   - (j^2{w}^2 +1) (1-2{w}) \nonumber\\
&=& 2\bar E +2{w} - j^2{w}^2+2j^2{w}^3\nonumber\\
&=& 2j^2(w-w_1)(w-w_2)(w-w_3)\,.
\eea
In a large $j$ expansion limit one finds 
\bea
w_1 &=& \frac{1}{2}-\frac{2(1+2\bar{E})}{j^2}-\frac{8(1+6\bar{E}+8 \bar{E}^2)}{j^4}\nonumber\\
&-&\frac{64(1+10 \bar{E}+30 \bar{E}^2+28 \bar{E}^3)}{j^6}+O\left(\frac{1}{j^8}\right)\,,\nonumber\\
w_2 &=&\frac{\sqrt{2\bar{E}}}{j}+\frac{1+2\bar{E}}{j^2}+\frac{20 \bar{E}^2+12 \bar{E}+1}{2 \sqrt{2} j^3 \sqrt{\bar{E}}}  \nonumber\\
&+&\frac{4 \left(8 \bar{E}^2+6 \bar{E}+1\right)}{j^4}\nonumber\\
&+&\frac{3696 \bar{E}^4+3360 \bar{E}^3+840 \bar{E}^2+40 \bar{E}-1}{16 \sqrt{2} j^5
   \bar{E}^{3/2}}\nonumber\\
   &+&\frac{32 \left(28 \bar{E}^3+30 \bar{E}^2+10 \bar{E}+1\right)}{j^6}+O\left(\frac{1}{j^7}\right)\,,\nonumber\\
w_3 &=& {-}\frac{\sqrt{2} \sqrt{\bar{E}}}{j}{+}\frac{2 \bar{E}{+}1}{j^2}{-}\frac{20 \bar{E}^2{+}12 \bar{E}{+}1}{2 \sqrt{2} j^3 \sqrt{\bar{E}}}\nonumber\\
&+&\frac{4 \left(8 \bar{E}^2+6 \bar{E}+1\right)}{j^4}\nonumber\\
&-&\frac{3696 \bar{E}^4+3360 \bar{E}^3+840 \bar{E}^2+40 \bar{E}-1}{16 \sqrt{2} j^5\bar{E}^{3/2}}+\nonumber\\
&+&\frac{32 \left(28 \bar{E}^3+30 \bar{E}^2+10 \bar{E}+1\right)}{j^6}\nonumber\\
&+& O\left(\frac{1}{j^7}\right)\,,
\eea
with $w_{\rm max}=w_1$ which we also express in terms of $\gamma$
\bea
w_{\rm max}&=& \frac{1}{2}-\frac{2 \gamma ^2}{j^2}+\frac{8 \gamma ^2(1-2\gamma^2)}{j^4}\nonumber\\
&-& \frac{32 \gamma^2 \left(7 \gamma ^4-6 \gamma ^2+1\right)}{j^6}+O\left(\frac{1}{j^8}\right)\,.
\eea
It is natural to introduce the new variable $ \xi  =\frac{w}{w_{\rm max}}$ and the small quantity $\epsilon$ defined as \cite{Bini:2017wfr} 
\beq
\epsilon=\frac{2}{j \sqrt{2\bar E}}\,,
\eeq
so that
\bea
{\xi} &=& \frac{w}{w_{\rm max}}=w\Big[2+4 \epsilon ^2 \bar{E} \left(2 \bar{E}+1\right)\nonumber\\
&+&16 \epsilon ^4 \bar{E}^2 \left(6 \bar{E}^2+5 \bar{E}+1\right)\nonumber\\
&+&16 \epsilon ^6 \bar{E}^3 \left(96 \bar{E}^3+112 \bar{E}^2+42 \bar{E}+5\right)\Big]\nonumber\\
&+& O(\epsilon^8)\,,
\eea  
where ${\xi}$ (now, in the large $j$ expansion limit)  varies from 0 to 1. In ${\xi}=1$  the integral is singular and one should consider its finite part.

The complete TS result  reads 
\bea
\label{GeodesicScattAng}
\frac{\chi(\alpha,\gamma,  j)+\pi}{2}&=& \sum_{k=1}^\infty \frac{\chi_k(\alpha,\gamma)}{j^k}\,,
\eea
where we used $\hat E^2-1\equiv \gamma^2-1=2\bar E$ (i.e., $\hat E=\gamma$) as the energy parameter. 
Here $\chi_k=\chi_k(\alpha,\gamma)$ with a specific structure at the various orders in $j$, 
and are summarized in Table \ref{tab:chi_alphas} and there are no limitations to extend the expansions lo larger orders in $1/j$. The limit $\alpha \to 0$ reproduces the well-known Schwarzschild result.

\begin{table*}  
\caption{\label{tab:chi_alphas} Coefficients of the expansion in powers of $\alpha=r_b/r_s$ of the scattering angle in a TS spacetime, from $\alpha^0$ (Schwarzschild) to $\alpha^4$.  Note that the $\alpha^n$ contributions starts by $j^{-n}$ implying that the final result is a large-$j$ expansion of the exact result whereas $\alpha$ does not need to be taken as small.
}
\begin{ruledtabular}
\begin{tabular}{ll}
$\chi_1$ & $\frac{2 \gamma ^2-1}{\sqrt{\gamma ^2-1}}+\alpha \sqrt{\gamma^2-1}$\\
$\chi_2$ & $\frac{3}{8} \pi  \alpha ^2 \left(\gamma ^2-1\right)+\frac{1}{4} \pi 
\alpha  \left(3 \gamma ^2-1\right)+\frac{3}{8} \pi  \left(5 \gamma
    ^2-1\right)$\\
$\chi_3$ & $\frac{5}{3} \alpha ^3 \left(\gamma ^2-1\right)^{3/2}+\alpha ^2
\sqrt{\gamma ^2-1} \left(4 \gamma ^2-1\right)+\frac{\alpha  \left(8
\gamma ^4-8
    \gamma ^2+1\right)}{\sqrt{\gamma ^2-1}}+\frac{64 \gamma ^6-120 \gamma
^4+60 \gamma ^2-5}{3 \left(\gamma ^2-1\right)^{3/2}}$\\
$\chi_4$ & $\frac{105}{128} \pi  \alpha ^4 \left(\gamma ^2-1\right)^2+\frac{15}{32}
\pi  \alpha ^3 \left(5 \gamma ^4-6 \gamma ^2+1\right)+\frac{9}{64} \pi
    \alpha ^2 \left(35 \gamma ^4-30 \gamma ^2+3\right)+\frac{15}{32} \pi 
\alpha  \left(21 \gamma ^4-14 \gamma ^2+1\right)+\frac{105}{128} \pi
    \left(33 \gamma ^4-18 \gamma ^2+1\right)$\\
$\chi_5$ & $\frac{21}{5} \alpha ^5 \left(\gamma ^2-1\right)^{5/2}+\frac{7}{3} \alpha
^4 \left(\gamma ^2-1\right)^{3/2} \left(6 \gamma ^2-1\right)+2 \alpha ^3
    \sqrt{\gamma ^2-1} \left(16 \gamma ^4-12 \gamma
^2+1\right)+\frac{\alpha ^2 \left(4 \gamma ^2 \left(3-4 \gamma
    ^2\right)^2-2\right)}{\sqrt{\gamma ^2-1}}$\\
&$+\frac{\alpha  \left(8
\left(48 \gamma ^6-112 \gamma ^4+84 \gamma ^2-21\right) \gamma
^2+7\right)}{3
    \left(\gamma ^2-1\right)^{3/2}}+\frac{630 \gamma ^2+32 \left(56
\gamma ^6-180 \gamma ^4+210 \gamma ^2-105\right) \gamma ^4-21}{5
\left(\gamma
    ^2-1\right)^{5/2}}$\\
$\chi_6$ & $\frac{1155}{512} \pi  \alpha ^6 \left(\gamma
^2-1\right)^3+\frac{315}{256} \pi  \alpha ^5 \left(\gamma ^2-1\right)^2
\left(7 \gamma
    ^2-1\right)+\frac{525}{512} \pi  \alpha ^4 \left(\gamma ^2-1\right)
\left(21 \gamma ^4-14 \gamma ^2+1\right)$\\
&$+\frac{25}{128} \pi  \alpha ^3
    \left(21 \gamma ^2 \left(11 \gamma ^4-15 \gamma
^2+5\right)-5\right)+\frac{105}{512} \pi  \alpha ^2 \left(429 \gamma
^6-495 \gamma ^4+135
    \gamma ^2-5\right)+\frac{315}{256} \pi  \alpha  \left(143 \gamma
^6-143 \gamma ^4+33 \gamma ^2-1\right)$\\
&$+\frac{1155}{512} \pi  \left(221
\gamma
    ^6-195 \gamma ^4+39 \gamma ^2-1\right)$\\
 \hline
\end{tabular}
\end{ruledtabular}
\end{table*}
An example of numerical study of the scattering angle (without self force corrections) is shown in Fig. \ref{fig_geoTS}.\\
 \begin{figure}
     \centering
     \includegraphics[scale=0.9]{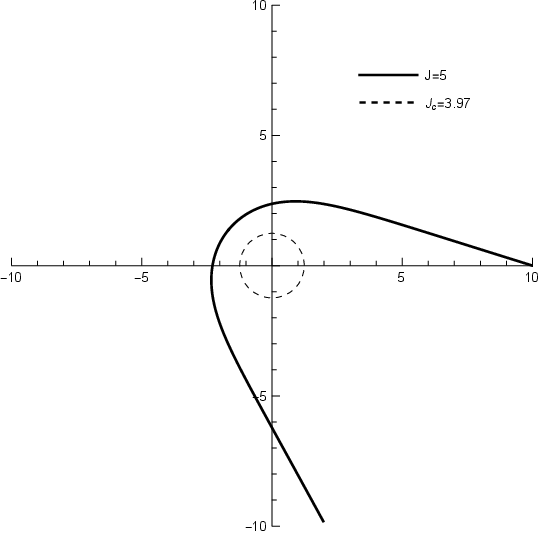} 
\caption{\label{fig_geoTS} Scattering orbit in a TS spacetime for the following choice of parameters:  $j=5$,
$\bar{E}=2$, $r_b={w}_b^{-1}=1$ and $r_s=0.8$.
With this choice of parameters, the critical radius (dashed circle) and the critical
angular momentum are $r_c=1.24$ and $j_c=3.97$.
}
\end{figure}

\section{Including scalar self force effects}
\label{scalSFeffectc}
Self force affects the particle's (geodesic) four velocity with $O(q^2)$ modifications.
Denoting the particle's four velocity as $u^\alpha(\tau)$ with $\tau$ the proper time parameter, the equations of motion read
\beq
\frac{Du^\alpha}{d\tau}=u^\mu \nabla_\mu u^\alpha=\frac{F^\alpha}{m}\equiv \hat F^\alpha(\tau)\,.
\eeq
Here, both $u$ and $F$ have components only along $t,r,\phi$ and 1) $u^\alpha$ is timelike and unitary:  $u\cdot u=-1$, while 2) $F$ is orthogonal to $u$:
$F\cdot u=0$. This implies
\beq
g_{tt}(u^t)^2+g_{rr}(u^r)^2+g_{\phi\phi}(u^\phi)^2+1=0\,,
\eeq
from which
\beq
\label{norm_cond}
(u^r)^2=-\frac{g_{tt}}{g_{rr}}(u^t)^2-\frac{g_{\phi\phi}}{g_{rr}}(u^\phi)^2-\frac{1}{g_{rr}}\,,
\eeq
and
\beq
u_tF^t+u_rF^r+u_\phi F^\phi=0\,,
\eeq
from which
\beq
F^r=-\frac{u_t}{u_r}F^t-\frac{u_\phi}{u_r} F^\phi\,.
\eeq
Note that the force terms only exist along the particle's orbit. Therefore,  the metric coefficients in Eq. \eqref{norm_cond} are meant to be restricted to the particle's world line.
The equations of motion yield the following coupled system of differential equations
\bea
\label{motion_eqs}
&&\frac{dt}{d\tau}=u^t\,,\quad \frac{dr}{d\tau}=u^r\,,\quad \frac{d\phi}{d\tau}=u^\phi\,,\nonumber\\
&& \frac{du^\alpha}{d\tau}+\Gamma^{\alpha}{}_{\mu\nu}u^\mu u^\nu =\hat F^\alpha\,,
\eea
where
\beq
\Gamma^{\alpha}{}_{\mu\nu}=\Gamma^{\alpha}{}_{\mu\nu}(x)\big|_{x=x(\tau)}
\eeq
are (known) functions of the coordinates, to be restricted to the particle's world line.
Eqs. \eqref{motion_eqs} are conveniently written in terms of covariant components.
More precisely,  
\beq
\label{eqs_cov}
u_\sigma\frac{d g^{\alpha\sigma}}{d\tau}+g^{\alpha\sigma}\frac{d u_\sigma}{d\tau}+ g^{\alpha\sigma}\left( g_{\sigma\mu,\nu} -\frac12 g_{\mu\nu,\sigma}\right)u^\mu u^\nu =g^{\alpha\sigma}\hat F_\sigma\,.
\eeq
Eq. \eqref{eqs_cov} for $\alpha=t$ becomes
\bea
\label{eqs_cov_t}
&& u_t \frac{d \ln g^{tt}}{d\tau}+\frac{d u_t}{d\tau}+  g_{tt,r}  u^t u^r = \hat F_t\,,\nonumber\\
&&\frac{d u_t}{d\tau}  = \hat F_t\,.
\eea
Similarly, for $\alpha=\phi$
\bea
\label{eqs_cov_phi}
&&\frac{d u_\phi}{d\tau}  = \hat F_\phi\,.
\eea
The equation for $u_r$, instead, is more involved
\bea
\label{eqs_cov_r}
&& u_r\frac{d \ln g^{rr}}{d\tau}+ \frac{d u_r}{d\tau}+ \left( g_{r\mu,\nu} -\frac12 g_{\mu\nu,r}\right)u^\mu u^\nu = \hat F_r\nonumber\\
&& \frac{d u_r}{d\tau} -\frac12 g_{\mu\nu,r}u^\mu u^\nu = \hat F_r\,,
\eea
and can be cast in the suggestive form
\beq
\frac{d u_r}{d\tau} =\hat F_r^{\rm g}+ \hat F_r\,,
\eeq
where   
\beq
\hat F_r^{\rm g}=\frac12 g_{\mu\nu,r}u^\mu u^\nu\,.
\eeq
Equivalently, Eq. \eqref{eqs_cov_r}   can be replaced by  the normalization condition \eqref{norm_cond}.

We can solve the equations of motion perturbatively. To this end let us write
\bea
t(\tau)&=&t_g(\tau)+\delta t(\tau)\,,\quad r(\tau)=r_g(\tau)+\delta r(\tau)\,,\nonumber\\ 
\phi(\tau)&=& \phi_g(\tau)+\delta \phi(\tau)\,,
\eea
where $\delta x^\alpha(\tau)$ denote deviations from geodesic values.
Consequently
\bea
u^\alpha&=& u_g^\alpha +\frac{d \delta x^\alpha}{d\tau}\,,\nonumber\\
u_\alpha&=& g_{\alpha\beta}\left( u_g^\beta +\frac{d \delta x^\beta}{d\tau}\right)\,.
\eea
Explicitly
\bea
u_t&=&-E -f_s  \frac{d \delta t}{d\tau} 
+ \frac{r_s}{r_g^2 } \frac{E}{f_s} \delta r  \nonumber\\
&\equiv & -E+\delta u_t\,,\nonumber\\
u_\phi &=&L +r_g^2 \frac{d \delta \phi}{d\tau} 
+\frac{2L}{r_g} \delta r  \nonumber\\
&\equiv & L+\delta u_\phi\,,\nonumber\\
u_r&=&\frac{1}{f_sf_b}\frac{dr_g}{d\tau}   
+\frac{1}{f_sf_b}\frac{d \delta r}{d\tau}\nonumber\\  
&+&
\frac{ (2 f_b f_s -f_s-f_b)}{r_g f_s^2 f_b^2}\frac{dr_g}{d\tau} \delta r \nonumber\\
&\equiv & \frac{1}{f_sf_b}\frac{dr_g}{d\tau}+\delta u_r\,,
\eea
where
\beq
f_{s,b}=1-\frac{r_{s,b}}{r_g(\tau)}\,,
\eeq
and
\beq
\partial_r g_{tt}|_{r=r_g}=\frac{r_s}{r_g^2}\,,\qquad
\partial_r g_{rr}|_{r=r_g}=\frac{(2 f_b f_s-f_s-f_b)}{r_g f_s^2 f_b^2}\,.
\eeq
Note that the quantities $\delta u_\alpha$ are {\it defined}  from the previous relations, namely
\bea
\delta u_t&=& -f_s  \frac{d \delta t}{d\tau} 
+ \frac{r_s}{r_g^2 } \frac{E}{f_s} \delta r \,, \nonumber\\
\delta u_\phi &=& r_g^2 \frac{d \delta \phi}{d\tau} 
+\frac{2L}{r_g} \delta r\,,  \nonumber\\
\delta u_r&=&  
 \frac{1}{f_sf_b}\frac{d \delta r}{d\tau}
+\frac{ (2 f_b f_s -f_s-f_b)}{r_g f_s^2 f_b^2}\frac{dr_g}{d\tau} \delta r \,,\qquad
\eea
in the sense that they are not obtained by lowering indices of the corresponding contravariant quantities $\delta u^\alpha$ (in fact, only the components of the full object $u^\alpha$ have a meaning when the indices raised and lowered with the metric).

Using the normalization condition \eqref{norm_cond} (instead of the radial equation which would give an equivalent but second order equation) we find the following first order equation for $\delta r$
\bea
\label{eq_delta_r_fin}
\frac{d \delta r}{d\tau}&=&  -\frac{f_b f_s L }{r_g^2\frac{dr_g}{d\tau}}\delta u_\phi 
-\frac{E f_b}{\frac{dr_g}{d\tau}}\delta u_t  \nonumber\\
&-& \frac12 \frac{(2 f_b f_s -f_s -f_b)}{r_g f_b f_s} \frac{dr_g}{d\tau}  \delta r \nonumber\\
&+& \frac12 \frac{(2 L^2 f_s^2+r_g r_s E^2) f_b}{ r_g^3 f_s \frac{dr_g}{d\tau}}  \delta r \,.
\eea
Here, from Eqs \eqref{eqs_cov_t} and \eqref{eqs_cov_phi}, we have also
\bea
\delta u_t(\tau)&=&\int_{-\infty}^\tau d\tau  \hat F_t\equiv -{\mathcal E}(\tau)\,,\nonumber\\
\delta u_\phi(\tau)&=&\int_{-\infty}^\tau d\tau  \hat F_\phi\equiv {\mathcal L}(\tau)\,,
\eea
where the duplicated notation $\delta u_t= -{\mathcal E}(\tau)$ and $\delta u_\phi= {\mathcal L}(\tau)$ is more standard in the literature and is a reminiscent of the physical meaning for these quantities  (energy and angular momentum).
It is convenient to introduce the compact notation
\bea
K_{\rm g}(\tau)&=&-\frac12 \frac{(2 f_b f_s -f_s -f_b)}{r_g f_b f_s} \frac{dr_g}{d\tau} \nonumber\\ 
&+& \frac12 \frac{(2 L^2 f_s^2+r_g r_s E^2) f_b}{ r_g^3 f_s \frac{dr_g}{d\tau}}\,,\nonumber\\
K_{\rm F}(\tau)&=& -\frac{f_b f_s L }{r_g^2\frac{dr_g}{d\tau}}\int_{-\infty}^\tau d\tau  \hat F_\phi\nonumber\\
&-& \frac{E f_b}{\frac{dr_g}{d\tau}}\int_{-\infty}^\tau d\tau  \hat F_t\nonumber\\
&=& -\frac{f_b f_s L }{r_g^2\frac{dr_g}{d\tau}}{\mathcal L}(\tau)+\frac{E f_b}{\frac{dr_g}{d\tau}}{\mathcal E}(\tau)\,,
\eea
so that Eq. \eqref{eq_delta_r_fin} becomes
\bea
\frac{d \delta r}{d\tau}&=& K_{\rm F}(\tau)+K_{\rm g}(\tau)\delta r (\tau)\,,
\eea
with solution
\beq
\delta r (\tau)=C_1 {\mathcal K}_{\rm g}(\tau)+{\mathcal K}_{\rm g}(\tau) \int_{-\infty}^\tau \frac{K_{\rm F}(\tau')}{{\mathcal K}_{\rm g}(\tau')}d\tau'\,,
\eeq
where $C_1$ is an integration constant  and we have defined
\beq
{\mathcal K}_{\rm g}(\tau)={\rm exp}\left(\int_{-\infty}^\tau K_{\rm g}(\tau') d\tau'\right)\,,
\eeq
or,  passing from $\tau$ to $t$ and then to $T= v t/b$, according to the relation
\beq
d\tau=\frac{dt}{u^t(t)}=\frac{b}{v}\frac{dT}{u^t(T)}\,,
\eeq
\beq
{\mathcal K}_{\rm g}(T)={\rm exp}\left(\frac{b}{v}\int_{-\infty}^T K_{\rm g}(T') \frac{dT'}{u^t(T')}\right)\,.
\eeq

The choice $C_1=0$ (equivalent to  $\lim_{\tau\to-\infty}\delta r(\tau)=0$) leads to the final form of the solution
\beq
\label{delta_r_fin}
\delta r (\tau)={\mathcal K}_{\rm g}(\tau) \int_{-\infty}^\tau \frac{K_{\rm F}(\tau')}{{\mathcal K}_{\rm g}(\tau')}d\tau'\,,
\eeq
or, in terms of the dimensionless variable $T$
\beq
\label{delta_r_finT}
\delta r (T)=\frac{b}{v}{\mathcal K}_{\rm g}(T) \int_{-\infty}^T \frac{K_{\rm F}(T')}{{\mathcal K}_{\rm g}(T')}\frac{dT'}{ u^t(T')}\,,
\eeq
The temporal and azimuthal components of the four velocity lead to the energy and angular momentum losses by the particle during the scattering process as functions of $\tau$
\bea
-u_t(\tau) 
&=& \hat E_-- \int_{-\infty}^\tau \hat F_t d\tau \equiv  \hat E_-+{\mathcal E}(\tau)\,, \nonumber\\
u_\phi(\tau)
&=& \hat L_-+ \int_{-\infty}^\tau \hat F_\phi d\tau \equiv   \hat L_-+ {\mathcal L}(\tau) \,,
\eea
where  $\hat E_-=E/(mc^2)$ and $ \hat L_-=L/m$ are the specific energy and angular momentum of the particle. Note that, in units of $c=1=G$,    $\hat E_-$ dimensionless and $j=\frac{L}{m c \frac{G M}{c^2} }=\frac{c}{GM}\hat L$ dimensionless.
Since $t$ and $\phi$ are associated with the Killing vectors of the background metric we will never need $t(\tau)$ and $\phi(\tau)$ to be  $O(q^2)$ corrected in order to evaluate the metric along the orbit of the source. 

Consequently, the background metric is not perturbed but the orbit of the particle is modified by the self force,
\beq
u^\phi(\tau)=\frac{u_\phi(\tau)}{r(\tau)^2\sin^2(\theta(\tau))}=\frac{u_\phi(\tau)}{r(\tau)^2}\,.
\eeq
The scattering angle $\chi$ is defined as
\beq
\chi+\pi =\int_{-\infty}^\infty  d\tau \, \frac{d\phi_p}{d\tau}
=\int_{-\infty}^\infty  d\tau \, u^\phi(\tau)\,,
 \eeq
and therefore 
\bea
\chi+\pi&=&\int_{-\infty}^\infty  d\tau \frac{\hat L_-+{\mathcal L}(\tau)}{[r_g(\tau)+\delta r(\tau)]^2}\nonumber\\
&=& \hat L_- \int_{-\infty}^\infty  \frac{d\tau}{[r_g(\tau)+\delta r(\tau)]^2}+
\int_{-\infty}^\infty  \frac{d\tau}{r_g^2(\tau)}{\mathcal L}(\tau) \nonumber\\
&=& \hat L_- \int_{-\infty}^\infty  \frac{d\tau}{ r_g^2(\tau)}\left(1-2 \frac{\delta r(\tau)}{r_g(\tau)}\right)\nonumber\\
&+&
\int_{-\infty}^\infty  \frac{d\tau}{r_g^2(\tau)}{\mathcal L}(\tau) \,.
\eea
We can distinguish the geodetic value
\bea
\chi_g+\pi
&=& \hat L_- \int_{-\infty}^\infty  \frac{d\tau}{ r_g(\tau)^2}\,,
\eea
with
\beq
\hat L_-= b \sqrt{\gamma^2-1}= b \gamma v \,,
\eeq
from the $O(q^2)$ self force corrected value
\bea
\delta \chi_{\rm SF}
&=& -2\hat L_- \int_{-\infty}^\infty  \frac{ \delta r(\tau)d\tau}{ r_g(\tau)^3} +
\int_{-\infty}^\infty  \frac{{\mathcal L}(\tau) d\tau}{r_g^2(\tau)}\nonumber\\
& \equiv & \delta \chi_{\rm SF}^{(r)}+\delta \chi_{\rm SF}^{(\phi)}\,.
\eea
Passing from $\tau$   to $T$   we find
\bea
\delta \chi_{\rm SF}^{(\phi)}&=&\frac{b^2}{v^2} \int_{-\infty}^\infty  \frac{ dT}{u^t(T) r_g^2(T)}\int_{-\infty}^T \frac{dT'}{u^t(T')}\hat F_\phi(T') \,,\qquad
\eea
and
\bea
\delta \chi_{\rm SF}^{(r)}&=&-2\hat L_- \int_{-\infty}^\infty  \frac{d\tau}{ r_g^3(\tau)} \delta r(\tau)\nonumber\\
&=& -2 b^2 \gamma  \int_{-\infty}^\infty  \frac{dT}{u^t(T) r_g^3(T)} \delta r(T)\,,\qquad
\eea
where $\delta r(T)$ is given in Eq. \eqref{delta_r_fin}.

\begin{table}  
\caption{\label{tab:notation}  Useful compact notation.
}
\begin{ruledtabular}
\begin{tabular}{ll}
${{f}}_n$ & $\frac{1}{(T^2+1)^n}$\\
$\tilde{{f}}_n$ & $\frac{T}{(T^2+1)^n}$\\
${{f}}_n^{\rm ask}$ & $\frac{{\rm arcsinh}^k(T)}{(T^2+1)^n}$\\
$\tilde f_n^{\rm ask}$ & $\frac{T{\rm arcsinh}^k(T)}{(T^2+1)^n}$\\
${{f}}^{\rm at0}$ & ${\rm arctan}(T)+\frac{\pi}{2}$\\
${{f}}_n^{\rm at1}$ & $\frac{{\rm arctan}(T)}{(T^2+1)^n}$\\
$\tilde{{f}}_n^{\rm at1}$ & $\frac{T{\rm arctan}(T)}{(T^2+1)^n}$\\
${{f}}_n^{\rm log}$ & $\frac{{\rm Log}(1+T^2)}{2(1+T^2)^n}$\\
 \hline
\end{tabular}
\end{ruledtabular}
\end{table}

Defining the dimensionless charge
\beq
\hat q=\frac{q}{\sqrt{r_s m}}\,,
\eeq
by using the results of Ref. \cite{DiRusso:2025lip} the structure of the self force components is the following
\bea
\hat F_t &=& \hat q^2 \left(\frac{r_s}{b}\right)^2 \frac{1}{b} \left\{ v^3[F_t^{(0,0)}(T)+vF_t^{(0,1)}(T)+\ldots ]\right.\nonumber\\
&+& \left.\frac{r_s}{b}[F_t^{(1,0)}(T)+vF_t^{(1,1)}(T)+\ldots ]\right\}\,,\nonumber\\
\hat F_r &=&  \hat q^2 \left(\frac{r_s}{b}\right)^2  \frac{1}{b}\left\{v[F_r^{(0,0)}(T)+vF_r^{(0,1)}(T)+\ldots]\right.\nonumber\\
&+& \left. \left(\frac{r_s}{b}\right)  \frac{1}{v} [F_r^{(1,0)}(T)+vF_r^{(1,1)}(T)+\ldots]\right\}\,,\nonumber\\
\hat F_\phi &=& \hat q^2 \left(\frac{r_s}{b}\right)^2 \left\{ v [F_\phi^{(0,0)}(T)+vF_\phi^{(0,1)}(T)+\ldots]\right.\nonumber\\
&+& \left. \left(\frac{r_s}{b}\right)  \frac{1}{v} [F_\phi^{(1,0)}(T)+vF_\phi^{(1,1)}(T)+\ldots]\right\}\,,
\eea
as well as
\bea
\frac{r_{\rm geo}(T)}{b}&=& \sqrt{1+T^2}+r_1(T)\epsilon+ r_2(T)\epsilon^2+O(\epsilon^3)\,,\nonumber\\
\phi(T)&=& {\rm arctan}(T) +\epsilon \phi_1+\epsilon^2 \phi_2+O(\epsilon^3) \,,\nonumber\\
u^t(T)&=& 1+u_1^{t}(T)\epsilon+u_2^{t}(T)\epsilon^2+O(\epsilon^3)\,,
\eea
where the various coefficients have an expansion in $v$ and are listed in Table \ref{tab:ri-phii-uti} below.

\begin{table*}  
\caption{\label{tab:ri-phii-uti} List of the various coefficients entering the PM expansion of the unbound geodesics in the TS spacetime, in powers of $\epsilon=\frac{r_s}{2bv^2}$.
Note that $r(0)=r_{\rm min}$, i.e., $T=0$ corresponds to the turning point, where the minimum approach distance is reached by the massive probe,
$r_{\rm min}=b \left[1-\epsilon +\frac12 \epsilon^2 (1-4v^2\eta^2) \right]+O(\epsilon^3)$.
}
\begin{ruledtabular}
\begin{tabular}{ll}
$\frac{r_1}{b}$ & $ -1+\tilde f^{\rm as1}_{1/2}-v^2(\alpha +3) \tilde f^{\rm as1}_{1/2}$\\
$\frac{r_2}{b}$ & $\frac{1}{2} f_{1/2}  +\tilde f^{\rm as1}_{1} 
+\frac12 f^{\rm as2}_{3/2}+v^2\Big[-2 f_{1/2}
-2(\alpha +3)\tilde f^{\rm as1}_{1} -(\alpha +3) f^{\rm as2}_{3/2} \Big]$\\
   $$ & $+v^4\Big[ -\frac{3}{2} \left(\alpha ^2+2 \alpha +5\right) \tilde f^{\rm at1}_{1/2}  +(\alpha +3)^2 \tilde f^{\rm as1}_1
+\frac12 (\alpha +3)^2 f^{\rm as2}_{3/2}\Big]$\\
 \hline
$\phi_1$ & $\tilde f_{1/2}+f^{\rm as1}_1 +v^2\Big[(\alpha +1)\tilde f_{1/2}  -(\alpha +3)f^{\rm as1}_1  \Big]$\\
$\phi_2$ & $2 f^{\rm as1}_{3/2}-\tilde f^{\rm as2}_2+v^2\Big[(\alpha +3) \tilde f_1 +2 (\alpha +3)\tilde f^{\rm as2}_2 
-2 (\alpha +4)f^{\rm as1}_{3/2}+(\alpha +3)\left(f^{\rm at0}-\frac{\pi}{2}\right)  \Big] $\\
&$v^4 \Big[\frac14 \left(3 \alpha ^2+2 \alpha +3\right) \tilde f_1 
-\frac{3}{2}\left(\alpha ^2+2 \alpha +5\right)f^{\rm at1}_1 
-(\alpha +3)^2 \tilde f^{\rm as2}_2  
+2 (\alpha +3) f^{\rm as1}_{3/2}+
\frac{1}{4} \left(3 \alpha ^2+2\alpha +3\right) \left(f^{\rm at0}-\frac{\pi}{2}\right)\Big]$\\
\hline
$u^t_1$ &$2 v^2f_{1/2}+v^4f_{1/2}$\\
$u^t_2$ & $ 2v^2(f_1-\tilde f^{\rm as1}_{3/2})+v^4(5f_1+(5+2\alpha))\tilde f^{\rm as1}_{3/2}$\\
\end{tabular}
\end{ruledtabular}
\end{table*}

Recalling the geodesic (unperturbed) values
\bea
u^t&=&\gamma_{{\rm PN},6}+2v^2 \gamma_{{\rm PN},4} \epsilon f_{1/2}\,,\nonumber\\
u^r&=& v \gamma_{{\rm PN},4} \tilde f_{1/2}\nonumber\\
&-&\epsilon v \left( 
\frac{ 4 \alpha v^2(v^2+2)+v^4+4 v^2-8 }{8} \tilde f_1 \right.\nonumber\\
&+&\left.\frac{ 4 \alpha v^2(v^2+2)+9 v^4+20 v^2-8}{8} f^{\rm as1}_{3/2}\right)\,,\nonumber\\
u^\phi&=&\frac{v}{b} \gamma_{{\rm PN},4} f_1\nonumber\\
&+&\frac{\epsilon v}{b} \left(\frac{ 4 \alpha v^2(v^2+2)+9 v^4+20 v^2-8}{4  }\tilde f_2^{\rm as1}\right.\nonumber\\
&+&\left. 2 \gamma_{{\rm PN},4} f_{3/2} \right)\,,
\eea
where the notation $\gamma_{{\rm PN},n}$, ${}[\gamma^{-1}]_{{\rm PN},n}$ stands for  $\gamma=1/\sqrt{1-v^2}$, $\gamma^{-1}$ PN expanded up to $O(v^n)$ included,  
we find
\bea
\frac{\delta r(T)}{\hat q^2   r_s}
&=& \epsilon v [\delta r^{(1,1)}(T)+v^2 \delta r^{(1,3)}(T)]\nonumber\\
&+& \epsilon ^2   v [\delta r^{(2,1)}(T)+v^2 \delta r^{(2,3)}(T)]+\ldots
\eea
with   
\bea
\delta r^{(1,1)}(T)&=&
-\frac13 f_{1/2} \,, \nonumber\\
\delta r^{(1,3)}(T)&=&  -\tilde f_1   -\frac{7}{30}f_{1/2}   -\frac23\alpha \tilde f_1   
-\frac{\alpha}{3}f_{1/2}\,,\nonumber\\
\delta r^{(2,1)}(T)&=& -\frac16 f^{\rm at0} \sqrt{1+T^2}+\frac{1}{3}+\frac16 \tilde f_{1/2}\nonumber\\
&-& \frac13 f_1  +\frac13 \tilde f^{\rm as1}_{3/2}\,,  \nonumber\\
\delta r^{(2,3)}(T)&=&    -\frac{1}{180}f^{\rm at0} (45\alpha-14)\sqrt{1+T^2}-\frac{1}{10}\nonumber\\
&+&  (\frac13 \alpha+\frac19)f^{\rm at1}_{1/2} -(\frac{1}{45}+\frac{1}{4}\alpha)\tilde f_{1/2} \nonumber\\ 
&+& \frac{1}{10}f_1 +\frac13 (3+2\alpha )f^{\rm as1}_1 \nonumber\\
&-&  2(\frac23 \alpha +1)\tilde f_{3/2}-\frac{23}{30}\tilde f^{\rm as1}_{3/2}  \nonumber\\
&-& \frac23(3+2\alpha) f^{\rm as1}_2\,.
\eea
We relegate   more accurate expansions for all these quantities to an associated supplemental material file.

\subsection{Conservative/Dissipative self force: components at $O(\epsilon)$}

The conservative sector of the forces at the leading PM (i.e., $O(\epsilon)$) order reads
\bea
\hat F_t^{\rm cons}&=&\frac{\hat{q}^2 r_s  \epsilon }{b^2}(\alpha +2) v^5 \tilde{{f}}_3  \,,\nonumber\\
\hat F_r^{\rm cons}&=&\frac{\hat{q}^2 r_s\epsilon}{b^2} (\alpha +2) \frac13 v^6\left(-2 {{f}}_{3/2}+16 {{f}}_{5/2}-17 {{f}}_{7/2}\right)\,,\nonumber\\
\hat F_\phi^{\rm cons}&=&\frac{\hat{q}^2 r_s\epsilon}{b}(\alpha +2) v^4  \left[\frac23 v^2 \tilde {{f}}_3  -\left(1+\frac{13}{6}   v^2\right)\tilde{{f}}_2 \right]\,,\nonumber\\
\eea
so that~\footnote{The relation \eqref{Delta_F_def} should not be confused with the orbital average of $F_\alpha$, given by
$$
\langle F_\alpha \rangle=\int_{-\infty}^\infty F_\alpha (\tau) d\tau=\frac{b}{v}\int_{-\infty}^\infty F_\alpha (T) \frac{dT}{u^t(T)}\,.
$$
}
\beq
\label{Delta_F_def}
\Delta F_\alpha=\int_{-\infty}^\infty F_\alpha (T) dT \,,
\eeq
gives
\bea
\Delta \hat F_t^{\rm cons}&=& 0\,,\nonumber\\
\Delta \hat F_r^{\rm cons}&=& 
-\frac{ 4 \hat q^2 v^6 r_s }{15 b^2} ( \alpha+2)\epsilon\,, \nonumber\\
\Delta \hat F_\phi^{\rm cons}&=& 0\,.
\eea
The dissipative sector of the forces again at the leading PM (i.e., $O(\epsilon)$) order  is given by
\begin{widetext}
\bea
\hat F_t^{\rm diss}&=&\frac{\hat{q}^2 r_s\epsilon}{b^2} v^4 \left(-\frac{20 v^2}{3}{{f}}_{7/2}+\frac{(-30 \alpha+121) v^2+30
   }{30}{{f}}_{5/2}+\frac{(10 \alpha +13) v^2-10}{15}{{f}}_{3/2}\right)\,,\nonumber\\
\hat F_r^{\rm diss}&=&\frac{\hat{q}^2 r_s \epsilon}{b^2} v^3 c \left(\frac{(6 \alpha +1) v^2}{3}\tilde{{f}}_3 +\frac{-(10 \alpha  +13)  
   v^2+10}{15}\tilde{{f}}_2\right)\,,\nonumber\\
\hat F_\phi^{\rm diss}&=&\frac{\hat{q}^2 r_s\epsilon}{b}v^3 \left( (2 \alpha +7) v^2{{f}}_{5/2}-\frac{(50 \alpha +157)
   v^2+10}{30}{{f}}_{3/2}\right)\,,
\eea
with
\bea
\Delta \hat F_t^{\rm diss}&=& \frac{\hat{q}^2 r_s\epsilon}{b^2} v^4 \left(-\frac{20 v^2}{3}\frac{16}{15}+\frac{(-30 \alpha+121) v^2+30
   }{30}\frac{4}{3}+\frac{(10 \alpha +13) v^2-10}{15}2\right)\,, \nonumber\\
\Delta \hat F_r^{\rm diss}&=& 0\,,\nonumber\\
\Delta \hat F_\phi^{\rm diss}&=&\frac{\hat{q}^2 r_s\epsilon}{b}v^3 \left( (2 \alpha +7) v^2\frac43 -\frac{(50 \alpha +157)
   v^2+10}{15 }\right)  \,.
\eea

\subsection{Scattering angle}

Using the results of Ref. \cite{DiRusso:2025lip} (see the associated supplemental material file), we have both a SF corrected conservative ($v-$even) and dissipative ($v-$odd) values  of the scattering angle $\delta \chi_{\rm SF}$ as summarized below
\bea
\label{SFcorrectScattAng}
\delta\chi^{\rm cons}&=&\hat q^2 \frac{r_s}{b} \left\{  
\epsilon \pi(2 + \alpha) \left(-\frac{1}{8}v^2   - \frac{17}{24}    v^4  + \frac{25}{192}    v^6  \right) \right.\nonumber\\
&+&  \epsilon^2 \left[
 -2(2 +\alpha) v^2  \right. \nonumber\\
&+& 
 v^4   \left(\frac{(2+\alpha)(-401 + 30\alpha)}{45}    + \frac{\pi^2}{8} (1 +\alpha) (5 + \alpha) 
-\frac{16}{3} (2 + \alpha) \ln (2v\sigma_\phi) \right)\nonumber\\ 
&+& \left.\left.
 v^6   \left(-\frac{(2 + \alpha) (-3369 + 6550\alpha)}{1350}  + \pi^2 \left(-\frac{67}{48} - \frac{119}{96} \alpha  
- \frac{23}{96} \alpha^2 \right)
         + \frac{16}{9}(2+\alpha) \ln (2v\sigma_\phi)\right)\right]  \right\}\,,\nonumber\\
\delta\chi^{\rm diss}&=& \epsilon^2 \hat q^2 \frac{r_s}{b}\left(
\frac{2 v }{3}  + \frac{v^3 (17 + 20 \alpha)}{15}  + \frac{ v^5}{120 }
  (-80 + 56 \alpha + 
      80 \alpha^2 + \pi^2 (3 + 20 \alpha))\right) \,.
\eea
In the conservative scattering angle, at $O(\epsilon^2,v^4)$,  the
dimensionless scale $\sigma_\phi$, which we used as a cut-off regulator
in the $\phi$-component of the self-force, appears.

Actually, this scale was left explicit in the force component on purpose, in
order  to trace  back its presence at any moment. This occurrence is
similar to what shown in Ref. \cite{Barack:2023oqp}  for the
Schwarzschild conservative scattering angle, with the difference that we
here see this kind of terms already at $O(\epsilon^2)$~\footnote{As already said, switching off the TS parameter $\alpha$ is not sufficient to reproduce the Schwarzschild situation~\cite{Barack:2023oqp,Bini:2024icd}, because of the different boundary conditions imposed in the Schwarzschild case (purely ingoing conditions at the horizon) and in the TS case (regularity at the cap). In other words, \lq\lq TS plus $\alpha\to 0$ is not Schwarzschild."}.
At the same level of accuracy, in the dissipative part of the scattering angle, such a scale does not appear. 
It is then \lq\lq a fact" that the regularization procedure as used here enters the complete (conservative plus dissipative) scattering angle.
Therefore,
if/when another approach will compute the same quantity, 
a proper choice
of $\sigma_\phi$ can be made to produce agreement. 
This circumstance by itself motivates future investigations, in the sense that 1) looking for different regularization procedures will hopefully offer more insights for what concerns the most suited mathematical approach to the problem; 2) the support of some numerical analysis is needed to reach more solid conclusions.
Moreover, 3) our study is limited to either low PM and PN orders, because of the additional computational difficulties added here by the TS parameter $\alpha$  with respect to the Schwarzschild case: exploring (even only numerically) higher PM-PN orders should be considered too; 4) high-frequency expansions results should be studied externally to the MST formalism, in view of possible subtleties~\footnote{We thank T. Damour for raising this issue.} when exchanging the PM and PN expansions as we do in general.
It is therefore clear that this analysis, besides enriching existing discussions in the black hole cases, also poses additional questions which we leave for the future.

\end{widetext}

\section{Discussion and concluding remarks}
\label{Conclusion}
Relying on our recent study of scalar self-force effects for unbound orbits in the background of a Topological Star \cite{DiRusso:2025lip}, we have computed the scattering angle (both the conservative and the dissipative parts) for a small mass neutral probe ($\mu \ll M$). Our main results are summarized in Eqs. \eqref{SFcorrectScattAng}. The addition of self-force terms corrects in a significant way the geodetic result in Eq. \eqref{GeodesicScattAng} with coefficients $\chi_k$ in Table \ref{tab:chi_alphas}. 

We have also characterized the electromagnetic source in terms of a Papapetrou Field compatible with the Killing structure  of the background and further characterized TSs in terms of their super-energy tensors, sectional curvature, and geometric transport, both `parallel' and Fermi-Walker. 

Extending the analysis to gravitational or (for charged probes) electromagnetic self-force effects looks doable in principle but rather involved in practice. On the other hand, scalar self-force effects in `rotating' TSs  might be achieved with lesser effort and will be addressed in future works.

\section*{Acknowledgments}

We thank E.~Barausse, G.~Bonelli, A.~Cipriani, F.~Fucito, A. Geralico, J.~F.~Morales, P.~Pani and D. Usseglio for useful discussions and comments. D. B.  acknowledges sponsorship of
the Italian Gruppo Nazionale per la Fisica Matematica
(GNFM) of the Istituto Nazionale di Alta Matematica
(INDAM). M. B. and G. D. R. thank the MIUR PRIN
contract 2020KR4KN2 ``String Theory as a bridge between Gauge Theories and Quantum Gravity'' and the
INFN project ST\&FI String Theory and Fundamental
Interactions for partial support.

\end{document}